\documentclass{pnasapp}

\usepackage{graphics,graphicx}
\usepackage{amssymb,amsfonts,amsmath}
\usepackage{pnastwoF}
\usepackage{color}

\definecolor{signitext}{cmyk}{1,.45,0,.18}
\definecolor{signiback}{cmyk}{.15,.07,0,.03}

\url{www.pnas.org/cgi/doi/10.1073/pnas.1403473111}
\copyrightyear{2014}
\issuedate{x}
\volume{y}
\issuenumber{z}
\footlineauthor{Robinson et al.}

\begin{document}
%

\title{Titan solar occultation observations reveal transit spectra of a hazy world}

\author{
    Tyler D. Robinson\affil{1}{NASA Ames Research Center, Moffett Field, CA 94035}\affil{2}{NASA Astrobiology Institute's Virtual Planetary Laboratory},
    Luca Maltagliati\affil{3}{Laboratoire d'\'{E}tudes Spatiales et d'Instrumentation en Astrophysique (LESIA), 
                                        Observatoire de Paris, CNRS, UPMC, Universit\'{e} Paris-Diderot, Meudon, France},
    Mark S. Marley\affil{1}{},
    \and
    Jonathan J. Fortney\affil{4}{Department of Astronomy and Astrophysics, University of California, Santa Cruz, CA 95064} }

%
\contributor{Accepted to Proceedings of the National Academy of Sciences of the United States of America}
\maketitle
\begin{article}
\begin{abstract} 
High altitude clouds and hazes are integral to understanding exoplanet observations, and 
are proposed to explain observed featureless transit spectra.  However, it is difficult to make 
inferences from these data because of the need to disentangle effects of gas absorption from 
haze extinction.  Here, we turn to the quintessential hazy world---Titan---to clarify how high 
altitude hazes influence transit spectra.  We use solar occultation observations of Titan's 
atmosphere from the Visual and Infrared Mapping Spectrometer (VIMS) aboard NASA's 
{\it Cassini} spacecraft to generate transit spectra.  Data span 0.88--5~$\mu$m at a resolution 
of 12--18~nm, with uncertainties typically smaller than 1\%.  Our approach exploits symmetry 
between occultations and transits, producing transit radius spectra that inherently include 
the effects of haze multiple scattering, refraction, and gas absorption.  We use a simple 
model of haze extinction to explore how Titan's haze affects its transit spectrum.  Our spectra 
show strong methane absorption features, and weaker features due to other gases.  Most 
importantly, the data demonstrate that high altitude hazes can severely limit the atmospheric 
depths probed by transit spectra, bounding observations to pressures smaller than 0.1--10~mbar, 
depending on wavelength.  Unlike the usual assumption made when modeling and interpreting 
transit observations of potentially hazy worlds, the slope set by haze in our spectra is not flat, 
and creates a variation in transit height whose magnitude is comparable to those from the 
strongest gaseous absorption features.  These findings have important consequences for 
interpreting future exoplanet observations, including those from NASA's {\it James Webb Space 
Telescope}.
\end{abstract}
%

%
%

%
\noindent\fcolorbox{signitext}{signiback}{\parbox{\dimexpr \linewidth-2\fboxsep-2\fboxrule}{%
\abstractfont \color{signitext} { Significance\\[5pt]} 
Hazes dramatically influence exoplanet observations by obscuring 
deeper atmospheric layers.  This effect is especially pronounced in transit spectroscopy, which 
probes an exoplanet's atmosphere as it crosses the disk of its host star.  However, exoplanet 
observations are typically noisy, which hinders our ability to disentangle haze effects from other 
processes.  Here, we turn to Titan, an extremely well-studied world with a hazy atmosphere, to 
better understand how high altitude hazes can impact exoplanet transit observations.  We use data 
from NASA's {\it Cassini} mission, which observed occultations of the Sun by Titan's atmosphere, 
to effectively view Titan in transit.  These new data challenge our understanding of how hazes 
influence exoplanet transit observations, and provide a means of testing proposed approaches for 
exoplanet characterization. }}
\vspace{4 mm}
%

\dropcap{C}louds and hazes are ubiquitous in the atmospheres of solar system 
worlds \cite{sanchezlavegaetal2004}.  Furthermore, it is now becoming apparent that 
high altitude hazes strongly influence observed spectra of exoplanets \cite{pontetal2008,
lecavelierdesetangsetal2008,singetal2009,beanetal2010,gibsonetal2011}.  These hazes 
can limit our ability to study the underlying atmosphere, especially in transit spectroscopy, 
where the opacity of an exoplanet's atmosphere is studied by observing the wavelength 
dependent dimming of the host star as the planet crosses the stellar disk 
\cite{seager&sasselov2000,charbonneauetal2002}.  Here, long path lengths through the 
atmosphere mean that even relatively tenuous haze layers can become optically thick 
\cite{fortney2005}.  Depending on the cloud or haze properties, the result can be a flat 
or smoothly varying spectrum which contains little information about the composition of 
the bulk of the exoplanet's atmosphere.

A major obstacle to interpreting observations of potentially hazy exoplanet 
atmospheres is a lack of understanding of how aerosols influence transit spectra.  A 
number of key physical processes are at play---gas absorption, atmospheric refraction, 
Rayleigh scattering, and multiple scattering by cloud and haze particles 
\cite{brown2001,hubbardetal2001}.  While models of atmospheric transmission effects 
on a transit exist \cite{fortneyetal2003,barman2007,millerriccietal2009,
kaltenegger&traub2009,dekok&stam2012}, the complexity and 
computational cost of implementing all of the aforementioned processes forces simplification 
of the problem.  As a result, models commonly treat clouds and hazes as an opaque, grey 
absorbing layer that prevents light from probing deeper levels.

Here, we turn to the archetypal hazy world---Titan---to shed light on how high altitude 
clouds and hazes can influence transit observations, thus developing a new bridge 
between Titan studies and exoplanetary science, where Titan analog worlds are currently 
being modeled  \cite{kemptonetal2012,howe&burrows2012,morleyetal2013} and may 
prove to be a very common class of planet in the Universe \cite{lunine2010}. 
Titan is ideally suited to this task, as it posses a haze that extends to pressures 
approaching $10^{-6}$~bar \cite{porcoetal2005,belluccietal2009}, and, unlike exoplanets, 
is extremely well-studied, including {\it in situ} observations \cite{brownetal2009}.  We link 
Titan to exoplanet transit observations using solar occultation observations, which have an 
analogous geometry to exoplanet transits, and have a long history of providing detailed 
information on the atmospheric composition and structure of Solar System worlds 
\cite{elliot&olkin1996,broadfootetal1979,hubbardetal1988}.

\begin{figure}
  \centerline{\includegraphics[scale=0.5, trim= 30mm 25mm 30mm 30mm, clip]{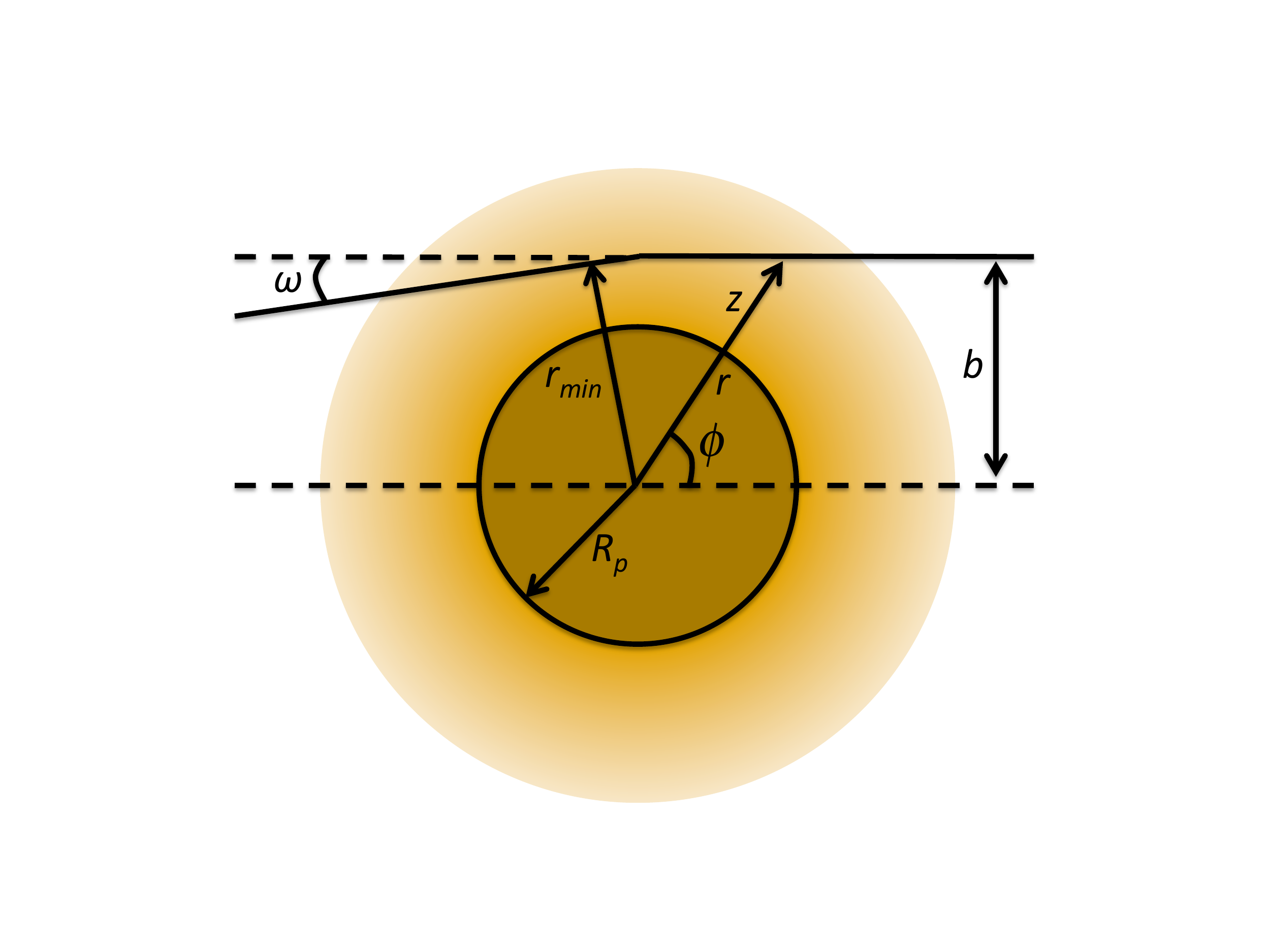}}

  \caption{Geometry and parameters relevant to occultation and transit.  In occultation, rays 
                enter from the right of the figure, with impact parameter $b$, are bent by atmospheric 
                refraction through an angle $\omega$, and have a distance of closest approach 
                $r_{min}$.  In transit, rays follow the opposite trajectory.  Note also the planetary 
                radius, $R_{p}$, the radial coordinate, $r$, the corresponding vertical height 
                coordinate, $z=r-R_{p}$, and the polar angle, $\phi$.}
  \label{fig:geom}
\end{figure}

To date, the similarities between exoplanet transits and solar or stellar occultations by solar 
systems worlds have not been exploited as a means of bridging these two research fields.  While 
Earth's atmospheric transmission has been measured during lunar eclipse and interpreted in 
terms of exoplanet observations \cite{palleetal2009,vidalmadjaretal2010}, these 
data only probe a limited range of altitudes ($\sim10$~km, depending on the solar elevation angle), 
and require corrections for telluric absorption, solar lines, and the lunar albedo.  Furthermore, it is 
clear that, with regards to exoplanetary science, what is needed is a better understanding of how 
high altitude hazes influence transmission spectra, and Earth does not have a particularly hazy 
upper atmosphere.

In this work, we use observations from NASA's {\it Cassini} mission of solar occultations by 
Titan's atmosphere to, for the first time ever, produce transit radius spectra of a hazy, 
well-characterized world.  Due to the symmetry in the geometry of occultations and transits, these 
data inherently include the effects of refraction and aerosol multiple scattering.  Our 
observations provide an essential and much-needed means of validating exoplanet transit models 
against Solar System data, and can be used to test proposed approaches for deciphering transit 
spectra. To better understand how Titan's high altitude haze affects the transit spectra, we develop an 
analytic model of haze extinction.  Finally, we interpret our spectra within the context of exoplanet 
observations, yielding new insight into the effects of hazes on transit observations.

\section{Observations and Data Processing}

Figure~1 
shows the geometry and relevant variables of occultation and 
transit, which are analogous to one another.  In transit, rays leave the stellar disk at the left 
of the diagram, are refracted and attenuated, and exit the atmosphere to travel to the observer, 
who is effectively an infinite distance away.  In occultation, rays from the occulted star 
come from the right of the diagram.  For a distant star, these rays are parallel, while for solar 
occultations rays can be non-parallel, depending on the angular size of the Sun. 
These rays are attenuated and refracted by the atmosphere before exiting in the direction of 
the observer (a relatively short distance $D$ away).  Thus, occultation measurements can 
be readily converted into transit radius spectra.

\begin{figure}
  \centerline{\includegraphics[scale=0.5]{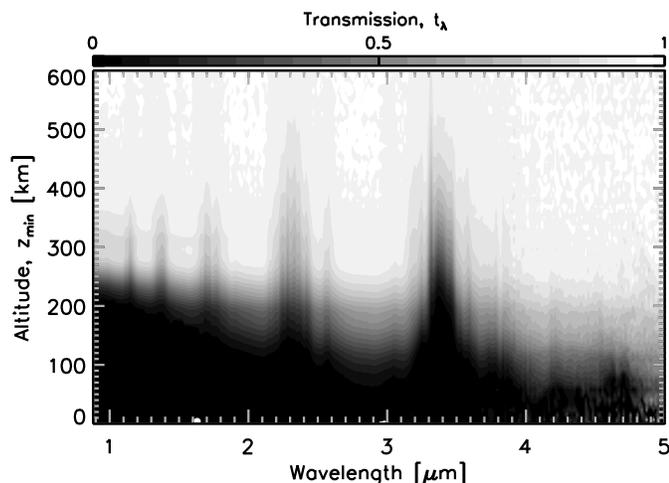}}

  \caption{Wavelength dependent transmission through Titan's atmosphere from the 
                27$^{\circ}$~N occultation.  The vertical axis is the ray's altitude of closest 
                approach, where an altitude of zero corresponds to the planetary surface, 
                at a radius of $R_{p}=2575$~km.  Darker shades indicate lower transmission, 
                and noise can be seen at transmission values very near to 1 and at wavelengths 
                beyond about 4.5~$\mu$m.}
  \label{fig:trans_contours}
\end{figure}

\subsection{Occultation Spectra}

The Visual and Infrared Mapping Spectrometer (VIMS) \cite{brownetal2004} aboard NASA's 
{\it Cassini} orbiter has observed ten solar occultations through Titan's atmosphere since the 
beginning of the mission. Spectra are acquired through a special solar port, which attenuates 
the intensity of sunlight on the detector, span 0.88--5~$\mu$m, and have a spectral resolution 
between 12--18 nm, increasing with wavelength.  Due to technical problems related to pointing 
stability and parasitic light, only four occultations out of ten could be analyzed. Table~1 
summarizes the main parameters of the four datasets.

\begin{table}[h]
  \centering
  {\bf Table 1. Parameters for Titan Solar Occultation Measurements} \\
  \begin{tabular}{r c c r c c}
             & {\it Cassini} &              &               & $D$  & Resolution  \\ 
    Date &      flyby       & Season & Latitude &  [km] &  [km] \\
    \hline
    Jan. 2006  & T10 &N winter &  70$^{\circ}$~S & 8300 & 15 \\
    Apr. 2009  & T53 & equinox &    1$^{\circ}$~N & 6300 & 7 \\
    Sep. 2011 & T78 & N spring &  40$^{\circ}$~N & 9700 & 10 \\
    Sep. 2011 & T78 & N spring &  27$^{\circ}$~N & 8400 & 10 \\
    \hline
  \end{tabular}
\end{table}

The atmospheric transmission, $t_{\lambda}$ along the line-of-sight is obtained by taking the ratio 
of every spectrum to the average spectrum outside the atmosphere (i.e., the reference solar 
spectrum). This is a self-calibrating method---instrumental effects and systematic errors are removed 
with the ratio, provided that the occultation is stable and the intensity variations are only due to the 
atmosphere. Figure~2 
shows the altitude-dependent transmission spectra for the 27$^{\circ}$~N occultation.

The uncertainty on the transmission values are given by the standard deviation over the 
average of the solar spectrum outside the atmosphere, which is stable except for random noise.  
Additional details on the data treatment process are described in Maltagliati et al. 
\cite{maltagliatietal2014}.  We note that these results are in good agreement with the analysis of 
the 70$^{\circ}$~S occultation dataset by Bellucci~et~al. \cite{belluccietal2009}, who employed 
different data processing methods.

Note that the angular diameter of the Sun at Saturn's orbital distance, $\theta_{\odot}$, is 
about 1~mrad, so that its image actually subtends a range of altitudes given by $\theta_{\odot} D$, 
or about 6--10~km.  Thus, each individual transmission spectrum contains information from a 
small range of altitudes.  Fortunately this range is smaller than both the vertical resolution of the 
corresponding datasets (shown in Table 1) and the atmospheric scale height ($\sim 40$~km, implying 
that atmospheric properties should not change dramatically over the 6--10~km range).  Nevertheless, 
future applications of the techniques described here may need to account for this ``smearing" effect, 
possibly by performing an analysis using a resolved portion of the solar disk (where, then, the 
relevant angular size is determined by the pixel or instrument field-of-view; see, e.g., 
\cite{gunsonetal1996,maltagliatietal2013}).

\subsection{Refraction Effects}

Refraction has two key effects on occultation observations. The first, and most familiar, is 
the bending of a light ray as it passes through the atmosphere. This effect is characterized 
by the refraction angle, $\omega$, which is the angle between the original ray path and the 
exit path.  Generally, the refraction angle is a function of wavelength (due to the 
wavelength dependent index of refraction of the atmosphere), and causes a distinction 
between a ray's impact parameter, $b$, and its distance of closest approach to the planet, 
$r_{min}$.  These parameters are all shown in Figure~1. 
Refraction is most pronounced for rays that pass near the surface, where molecular number 
densities are large.  Note, however, that, for Titan, strong attenuation by atmospheric haze 
particles at visible and near-infrared wavelengths largely limits sensitivity to the deep portions 
of the atmosphere where refractive bending of light rays is most significant.

The second key refractive effect is an apparent brightness loss, which is present even in 
the absence of molecular and aerosol attenuation \cite{elliot&olkin1996,hubbardetal1988}.  
This loss can be thought of as an apparent shrinking of the solar/stellar disk in the vertical 
direction or, equivalently, a spreading of rays from the source \cite{baum&code1953}. Here, 
brightness is diminished by a wavelength dependent factor $f_{\text{ref}}$, which is given 
by 
\begin{equation}
f_{\text{ref}} = \frac{1}{1 + D d\omega/dr_{\text{min}}} \ .
\end{equation}

To model these two effects, we use a ray tracing scheme described by 
van~der~Werf~\cite{vanderwerf2008}, which concisely outlines an accurate, fourth-order 
Runge-Kutta integration algorithm for tracking rays through an atmosphere.  The primary 
inputs to this model are profiles of atmospheric density and composition, as well as the 
refractive indexes of the major atmospheric constituents (which are, generally, 
wavelength dependent).  For Titan, we elect to use standard model profiles of atmospheric 
molecular number density and composition \cite{waiteetal2013}, as localized structure in 
measured profiles can lead to spurious features in our refraction calculations.  Our refraction 
models only include molecular nitrogen and methane in our computations, as these are the 
only major atmospheric constituents.  Finally, we use a measured, wavelength dependent 
refractivity for molecular nitrogen \cite{washburn1930} and a refractive index for methane 
of $1.0004478$ \cite{weber2002}, although our calculations are largely insensitive to this 
value due to the low mixing ratio of methane in the atmosphere.

By tracing rays on a fine grid of impact parameters (1 km vertical resolution from 
0--1500~km), we determine the relations between the impact parameter, altitudes 
of closest approach ($z_{\text{min}}=r_{\text{min}}-R_{p}$), refraction angle, and the refractive 
loss factor $f_{\text{ref}}$.  Our computed values are only weak functions of wavelength, as 
the refractivity of molecular nitrogen changes by less than 1\% over the wavelength range 
of interest.  Figure~3 
shows profiles of these parameters as a function of their altitude of closest approach, and 
demonstrates that, for our purposes, refraction effects are only important in the lowest 
100~km of the atmosphere.

\begin{figure}
  \centerline{\includegraphics[scale=0.5]{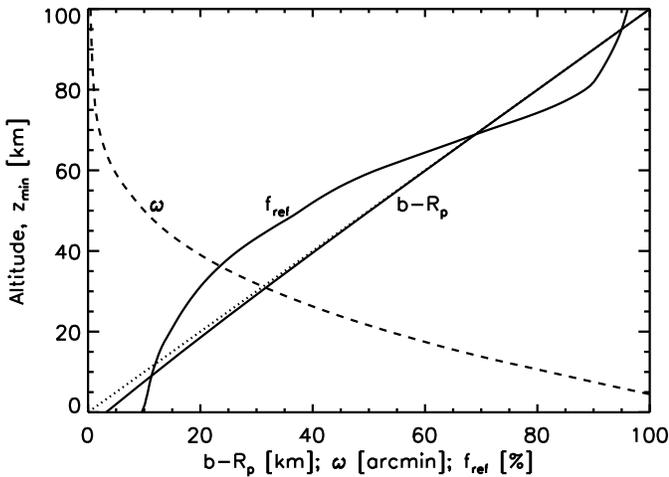}}

  \caption{Profiles of the impact parameter, $b$, refraction angle, $\omega$, and the refractive 
                loss factor, $f_{\text{ref}}$, from our Titan ray tracing model.  The vertical coordinate is the ray's 
                altitude of closest approach, $z_{\text{min}}$.  Parameters are shown for a wavelength 
                of 5~$\mu$m, where Titan's atmospheric haze is least opaque, and we use a distance, $D$, 
                from the spacecraft to Titan of 8400~km, appropriate for the 27$^{\circ}$~N occultation, when 
                computing the refractive loss.  The dotted line plots along the diagonal, and shows that the 
                impact parameter is always larger than the distance of closest approach.}
  \label{fig:refract_props}
\end{figure}
\begin{figure*}
  \centerline{\includegraphics[scale=0.6]{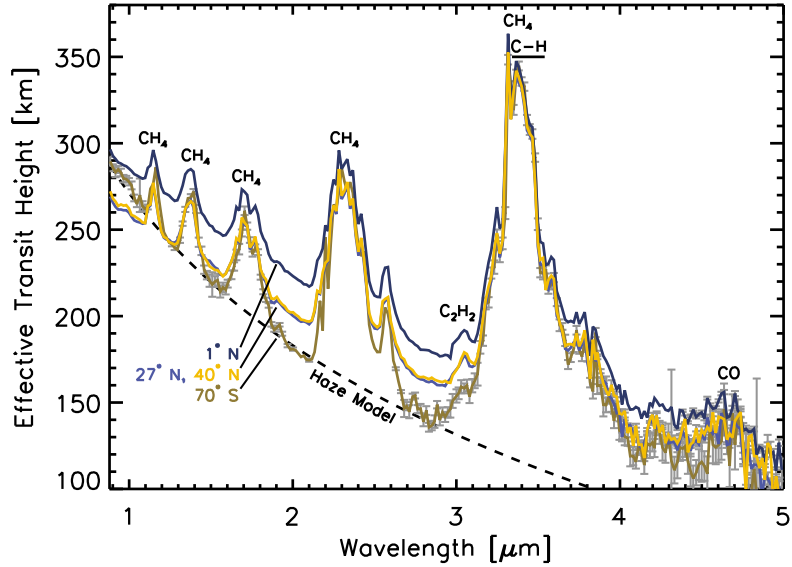}}

  \caption{Spectra of effective transit height, $z_{\text{eff},\lambda}=R_{\text{eff},\lambda}-R_{p}$,  for 
                all four {\it Cassini}/VIMS occultation datasets.  Key absorption features are labeled, and error 
                bars are shown only where the 1-$\sigma$ uncertainty is larger than 1\%.  Our best-fit haze model 
                for the 70$^{\circ}$S dataset is shown (dashed line).}
  \label{fig:transheight_all}
\end{figure*}

We note that refraction can also influence exoplanet transit observations under conditions where 
atmospheric opacity does not preclude light rays from reaching the deeper regions of an atmosphere.  
Here, the finite size of the host star paired with the geometry of refraction may prevent rays from probing 
altitudes below some critical height in the lower atmosphere \cite{betremieux&kaltenegger2013b}.  
Additionally, the transit signal may increase or decrease slightly due to the competing effects 
of refraction bending rays perpendicular to the limb while also focusing rays from within the planet's 
shadow towards the observer \cite{french1977,hubbard1977,hui&seager2002}.

\subsection{Computing Transit Spectra}

We define the transmission corrected for refractive losses at an impact parameter $b_{i}$ as 
$t^{\prime}_{\lambda,i} = t_{\lambda,i}/f_{\text{ref},i}$ (where a subscript `$i$' references the 
vertical gridding of the observed transmission spectra).  These can be converted into a transit 
radius spectrum by considering the attenuation produced by concentric annuli above Titan's 
surface.  An annulus has thickness $\pi \left( b_{i+1}^{2}-b_{i}^{2} \right)$, and we can define 
an effective transit radius as \cite{betremieux&kaltenegger2013}
\begin{equation}
R_{\text{eff},\lambda}^{2} = R_{\text{top}}^{2} - 
                                            \sum_{i=1}^{N} \left( \frac{t^{\prime}_{\lambda,i+1}+t^{\prime}_{\lambda,i}}{2} \right) 
                                                                     \left(  b_{\lambda,i+1}^{2}-b_{\lambda,i}^{2} \right) \ ,
\end{equation}
where $R_{\text{top}}=R_{p}+z_{\text{atm}}$ is the radial distance to the top of the atmosphere, 
whose altitude ($z_{\text{atm}}$) is large enough that atmospheric extinction and refraction are 
assumed to be negligible.  We also define an effective transit height as 
$z_{\text{eff},\lambda}=R_{\text{eff},\lambda} - R_{p}$, which is useful for identifying where in the 
atmosphere a given wavelength is probing.  Finally, note that the transit depth is proportional 
to $R_{\text{eff},\lambda}^{2}$, or, equivalently, $\left( z_{\text{eff},\lambda} + R_{p} \right)^{2}$.

\section{Transit Radius Spectra}

Figure~\ref{fig:transheight_all} shows the effective transit height, $z_{\text{eff},\lambda}$, for 
all four occultation datasets.  Error bars (1-$\sigma$) are shown where the errors are larger than 
1\% of the transit height, and key absorption features are identified.  In general, errors tend to be 
large beyond about 4~$\mu$m, where the solar flux is relatively weak.

The most obvious features are the methane bands at 1.2, 1.4, 1.7, 2.3, and 3.3~$\mu$m.  Weak 
absorption due to acetylene (C$_{2}$H$_{2}$) can be seen near 3.1~$\mu$m.  The 3.3~$\mu$m 
methane band is blended with other features, including the C-H stretching mode of aliphatic 
hydrocarbon chains appears near 3.4~$\mu$m \cite{belluccietal2009,maltagliatietal2014}.  An 
absorption feature of carbon monoxide, which forms from oxygen ions that precipitate into Titan's 
upper atmosphere and react with hydrocarbon species \cite{horstetal2008}, appears near 
4.6~$\mu$m, although data are particularly noisy here.  Finally, additional absorption has been 
noted in the 2.3 and 3.3~$\mu$m methane bands \cite{belluccietal2009,maltagliatietal2014}, 
which is due to other yet unknown species.

What is possibly the most interesting aspect of these spectra is the wavelength dependent slope 
of the continuum between the methane bands.  When observed across the full wavelength range, 
this slope produces a transit height variation that is comparable to, or larger than, the gaseous 
absorption features.  Assuming that the continuum is set by haze extinction, which shall be argued 
later, then the differences between the continuum levels for the four different datasets are related to 
different haze distributions (both vertically and in particle size) at the different latitudes/times of 
observation.  This is consistent with Titan's known hemispherical asymmetry 
\cite{sromovskyetal1981}, which may be caused by seasonally-varying atmospheric circulation 
patterns \cite{rannouetal2002}.  Note that methane clouds in Titan's atmosphere are found below 
about 30~km altitude \cite{rannouetal2006}, and do not affect our transit spectra, which probe much 
higher altitudes.

Finally, Figure~5 
shows an ``average" transit spectrum for Titan, in 
effective transit height and, as an example, in the transit depth signal for Titan crossing the solar disk.  
To obtain this result, we performed a weighted average of the four individual spectra in 
Figure~\ref{fig:transheight_all}.  The weights were determined from the latitude distribution of the 
individual spectra, assuming that the 70$^{\circ}$S spectrum is representative of latitudes 
between the south pole and midway to the 1$^{\circ}$N spectrum, that the 1$^{\circ}$N spectrum 
is representative of latitudes midway between the 70$^{\circ}$S spectrum and the 
27$^{\circ}$N spectrum, and so on.  Using these weights, we combine the spectra in 
$R_{\text{eff},\lambda}^{2}$, which is proportional to the transit depth signal.  While this 
weighted averaging is somewhat crude, as it ignores variations in longitude and time, the goal is 
only to produce a characteristic spectrum.

Standard deviations computed by comparing our characteristic spectrum to the individual spectra 
are shown as a shaded swath in Figure~5. 
The deviations are small 
near the peaks of methane bands, which probe higher in the atmosphere and are less sensitive 
to variations in the haze continuum levels.  At wavelengths dominated by haze opacity, though, the 
deviation is much larger.  Clearly future occultation measurements could help to better understand 
latitudinal and seasonal effects on our transit spectra, thus improving our characteristic spectrum.

\section{A Simple Haze Extinction Model}
To investigate the source and behavior of the continuum in our transit height spectra, we 
derived an analytic model of extinction by an opacity source that is distributed vertically in 
the atmosphere with scale height $H_{a}$, and whose absorption cross section, 
$\sigma_{\lambda}$ varies according to a power law in wavelength, with 
$\sigma_{\lambda} \propto \lambda^{\beta}$.  Ignoring refraction effects, which are negligible 
at most altitudes probed by our spectra, the wavelength dependent optical depth through 
the atmosphere for a given impact parameter is (see Appendix)
\begin{equation} \label{eqn:taulam}
\tau_{\lambda} = 2 \tau_{0} \left(\frac{\lambda}{\lambda_{0}}\right)^{\beta} \frac{b}{H_{a}} 
                           K_{1} \!\! \left( \frac{b}{H_{a}} \right) e^{(R_{p}+z_{0})/H_{a}} ,
\end{equation}
where $\tau_{0}$ is a reference optical depth at altitude $z_{0}$, and $K_{n}(x)$ is a 
modified Bessel function of the second kind.  With this model, a transit spectrum can be 
generated by finding the value of the impact parameter where $\tau_{\lambda} \approx 1$, 
which requires solving a transcendental expression.

We fit our analytic model to the continua in the 70$^{\circ}$S spectrum (selected since this 
dataset has been previously analyzed), and in the characteristic spectrum, which are shown 
in Figures~4 and 5, 
respectively.  The free 
parameters in this fit are the haze scale height, $H_{a}$, the reference optical depth, 
$\tau_0$, and the exponent in the cross section power law, $\beta$.  For the 70$^{\circ}$S 
dataset, we find $H_{a}=58 \pm 7$~km, $\tau_{0}=0.9 \pm 0.2$ (at 
$\lambda_{0}$=0.5~$\mu$m and $z_{0}=$200~km, which we will use hereafter), and 
$\beta=-2.2 \pm 0.2$.  For the characteristic spectrum, we find $H_{a}=55 \pm 8$~km, 
$\tau_{0}=0.8 \pm 0.4$, and $\beta=-1.9 \pm 0.2$.  Note that the slope of our power law is 
not due to pure Rayleigh scattering, which would have $\beta=-4$.  Instead it is due to the 
complexities of haze particle scattering between the limits of pure Rayleigh scattering and 
geometric optics.

Our parameters are in excellent agreement with {\it in situ} measurements reported by 
Tomasko~et~al. \cite{tomaskoetal2008}, who found $H_{a}=65$~km (with an uncertainty of 
20~km), $\tau_{0}=0.76$, and $\beta=-2.33$ above 80~km altitude .  For further comparison, 
Bellucci~et~al.~\cite{belluccietal2009}, in their analysis of the 70$^{\circ}$S occultation, 
found $H_{a}=$55--79~km, $\beta=-$1.7--2.2 between 120--300~km altitude, and 
$\tau_{0} \sim 0.6$.  Finally, Hubbard~et~al. \cite{hubbardetal1993}, in their analysis of 
stellar occultations by Titan's atmosphere, found $\beta=-1.7 \pm 0.2$.  These comparisons 
strongly support our conclusion that the continuum level in our transit spectra is set by Titan's 
high altitude haze.

\section{Implications}

The transit spectra shown in Figures~4~and~5 
demonstrate that high altitude hazes could have complex and important effects on exoplanet 
observations.  Note that our data span wavelengths that are nearly identical to (or larger than) 
the spectral coverage of the Near InfraRed Camera (NIRCam), Near InfraRed Spectrograph 
(NIRSpec), and the Near InfraRed Imager and Slitless Spectrograph (NIRISS) instruments that 
will launch aboard NASA's {\it James Webb Space Telescope}.  Thus, the spectra presented 
here indicate the types of haze effects that this mission may observe for transiting exoplanets.

For Titan, the haze continuum slope is strongly wavelength dependent, and is certainly not flat.  
This is contrary to what is commonly assumed in simple transit spectra models.  Clearly this 
continuum slope is of first order importance, as the magnitude of the transit height variations 
caused by the haze continuum is just as large as the observed gaseous absorption features.

Our transit spectra also show that haze opacity obscures information from the deep atmosphere, 
limiting the pressures probed to above $\sim0.1$~mbar at the shortest wavelengths, and 
$\sim10$~mbar at the longest wavelengths.  Even at the longest wavelengths, the altitudes 
probed are still 2--3 pressure scale heights above the surface.  Furthermore, at most continuum 
wavelengths in our spectra, haze limits sensitivity to pressures lower than (i.e., altitudes above) 
the $\sim1$~mbar level, with this effect becoming more severe at shorter wavelengths.  Thus, 
it is empirically possible for high altitude hazes to strongly limit the planetary characteristics that 
can be inferred from transit spectra, despite what others have claimed \cite{dewit&seager2013}.  
To further clarify this issue, it would be a very useful exercise to challenge current exoplanet 
retrieval models \cite{madhusudhan&seager2009,leeetal2012,benneke&seager2012,lineetal2013} 
with our Titan transit spectra, with the goal of improving our ability to understand and interpret 
transit observations of hazy exoplanets.

Looking to wavelengths beyond those analyzed here, we note that haze opacity effects in transit 
will become negligible in the mid-infrared, where refraction and gas absorption will then play a key 
role in limiting sensitivity to the lower atmosphere.  However, haze extinction will have much more 
dramatic  effects at ultraviolet and visible wavelengths, where Titan's haze particles are strongly 
absorbing \cite{tomaskoetal2008}.  This will make Rayleigh scattering effects undetectable---a ray 
passing through the atmosphere with a tangent height of $\sim300$~km (which is optimistic, as this 
is appropriate for the shortest wavelengths discussed here, not ultraviolet/visible wavelengths) 
will encounter $10^{20}$~molecules~cm$^{-2}$, which isn't optically thick to Rayleigh scattering by 
molecular nitrogen except at extreme ultraviolet wavelengths ($\sim40$~nm) and shorter.  Thus, 
Rayleigh scattering slopes in transit spectra, which have been proposed for constraining partial 
pressures due to spectrally inactive gases \cite{benneke&seager2012}, may not be accessible in 
hazy atmospheres.

\begin{figure}
  \centerline{\includegraphics[scale=0.5]{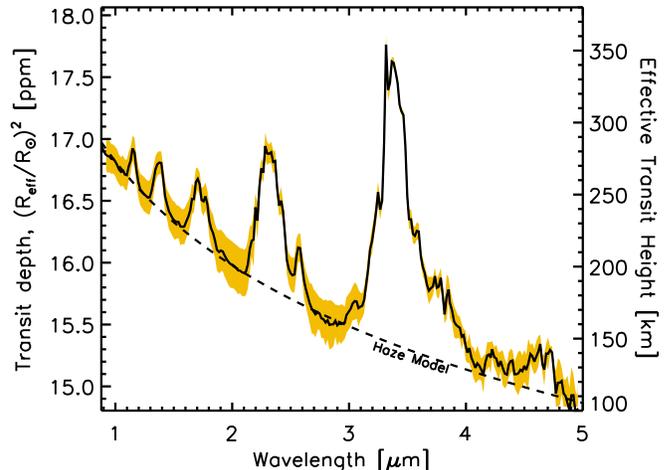}}

  \caption{Characteristic transit spectrum for Titan showing both the signal for Titan transiting the Sun 
               (left y-axis) and the effective transit height (right y-axis), assembled as a weighted mean 
               of the four spectra in Figure~\ref{fig:transheight_all}.  The shaded region indicates uncertainty 
               in our averaging, and is due to deviations from the mean in the four individual transit 
               spectra.  A best-fit haze model is shown (dashed).}
  \label{fig:transheight_combined}
\end{figure}

Recently, the 6 Earth mass transiting planet GJ~1214b \cite{charbonneauetal2009} has been the 
target of many observational campaigns to characterize the nature of its atmosphere 
\cite{beanetal2010,desertetal2011,bertaetal2012}.  This is the smallest planet with transit spectra 
observations, which appear to be flat at the 30 ppm level from 1.1--1.7~$\mu$m 
\cite{kreidbergetal2014}, and this trend may extend to 5~$\mu$m \cite{fraineetal2013}.  A Titan-like 
haze has been proposed as a viable explanation for these observations 
\cite{kemptonetal2012,howe&burrows2012,morleyetal2013}, and the data constrain this haze to be 
above the $10^{-1}$--$10^{-2}$~mbar level, depending on atmospheric composition 
\cite{kreidbergetal2014}.  While the methane concentration in the atmosphere of GJ~1214b is 
unknown, the high altitude haze interpretation isn't entirely consistent with the observations presented 
here.  Extending Titan's haze to the aforementioned low pressures would mask the methane 
features, but still would not produce a flat spectrum due to the wavelength-dependent haze opacity.  
Thus, a Titan-like haze on GJ~1214b would need to contain a continuum of effective particle 
radii that extends to sizes larger than is seen for Titan (whose haze particles have a characteristic 
size of about 1--2~$\mu$m \cite{tomaskoetal2008}, and are aggregates of smaller-sized 
monomers), as larger particles would tend to produce a flatter spectrum.  However, these larger-sized 
particles may be rather difficult to keep aloft at such low pressures \cite{spiegeletal2009}, especially 
given that the gravitational acceleration for GJ~1214b is nearly an order of magnitude larger than 
that in Titan's upper atmosphere.

\section{Conclusions}
We developed a technique for adapting occultation measurements of solar system worlds 
into transit radius spectra suitable for model validation and comparison to exoplanet observations.  
We applied this technique to Titan, deriving realistic spectra that inherently include effects due to 
gas absorption, refraction, and haze scattering, and used these spectra to better understand the 
effects of high altitude hazes on transit observations.  Absorption features due to methane are 
clearly visible, and weaker features due to acetylene, carbon monoxide, and a C-H stretching 
mode of aliphatic hydrocarbon chains. 

The continuum level in our spectra is set by Titan's extensive haze, and is well reproduced by an 
analytic haze extinction model derived here.  Haze has a dramatic effect on the transit spectra, 
limiting sensitivity to pressures smaller than 0.1--10~mbar, depending on wavelength.  Extinction 
from the haze imparts a distinct slope on the transit radius spectra, whose magnitude is comparable 
to that of the strongest gaseous absorption bands.  Thus, haze substantially impacts the amount of 
information that can be gleaned from transit spectra.

We note that the techniques used here apply equally well to occultation observations taken 
from orbit around any world.  Thus, there are opportunities empirically study the tenuous, dusty 
atmosphere of Mars \cite{maltagliatietal2013} and the atmosphere of Saturn \cite{banfieldetal2011} 
in the context of exoplanet transit spectroscopy.  Of course, numerous occultation observations 
exist for Earth \cite{gunsonetal1996}, which could be used to derive a transit spectrum of the only 
known habitable planet.  Finally, our understanding of how hazes influence transit spectra of Titan 
could be greatly improved by acquiring additional occultation observations in a {\it Cassini} extended 
mission.

\appendix
Given the extinction coefficient, $\alpha_{\lambda}=n_{a}\sigma_{\lambda}$, where 
$n_{a}$ is the absorber number density and $\sigma_{\lambda}$ is the 
wavelength dependent absorption cross section, the optical depth is determined by 
the integral
\begin{equation}
\tau_{\lambda} = \int n_{a}\sigma_{\lambda} ds ,
\end{equation}
where integration proceeds along a ray's path shown in Figure~1. 
Ignoring refraction, we have $ds = \left( R_{p} + z \right) d\phi/ \sin(\phi)$ and 
$R_{p} + z = b/\sin(\phi) $, so that
\begin{equation}
\tau_{\lambda} = 2 \int_{0}^{\pi/2} \frac{b}{\sin^{2}(\phi)}n_{a}\sigma_{\lambda} d\phi ,
\end{equation}
where we have exploited the symmetry about $\phi=\pi/2$.  If the absorber is 
distributed with a scale height $H_{a}$, with $n_{a} = n_{a0}\exp[-(z-z_{0})/H_{a}]$, where 
$n_{a0}$ is the number density at the altitude $z_{0}$, and assuming that the 
absorption cross section is a power law in wavelength, 
$\sigma_{\lambda} = \sigma_{\lambda0} \left(\lambda/\lambda_{0}\right)^{\beta}$, 
where $\lambda$ is wavelength, $\sigma_{\lambda0}$ is the fiducial value at $\lambda_{0}$, 
and $\beta$ defines the slope of the power law, we then have
\begin{equation}
\tau_{\lambda} = 2 \tau_{0} \left(\frac{\lambda}{\lambda_{0}}\right)^{\beta} 
                           \frac{b}{H_{a}} e^{\frac{R_{p}+z_{0}}{H_{a}}} \int_{0}^{\pi/2} \frac{e^{-b/H_{a}\sin{\phi}}}{\sin^{2}(\phi)} d\phi ,
\end{equation}
where $\tau_{0} = n_{a0}\sigma_{\lambda0}H_{a}$ is a reference vertical optical 
depth.  Making the substitution $\cosh (y) = 1/\sin(\phi)$, we have 
\begin{equation}
\tau_{\lambda} = 2 \tau_{0} \left(\frac{\lambda}{\lambda_{0}}\right)^{\beta} 
                           \frac{b}{H_{a}} e^{\frac{R_{p}+z_{0}}{H_{a}}} \int_{0}^{\infty} \cosh(y) e^{-\frac{b \cosh(y)}{H_{a}}} dy ,
\end{equation}
which has the analytic solution given in the main text (Equation~\ref{eqn:taulam}).  Note 
that, for large $b/H_{a}$, we have 
$K_{1}(b/H_{a}) \sim \sqrt{\pi/2} \exp(-b/H_{a})/\sqrt{b/H_{a}}$, so that 
Equation~\ref{eqn:taulam} gives 
\begin{equation}
\tau_{\lambda} \sim \tau_{0} \left(\frac{\lambda}{\lambda_{0}}\right)^{\beta} \sqrt{\frac{2\pi b}{H_{a}}} e^{-(b-R_{p}-z_{0})/H_{a}} ,
\end{equation}
which is in agreement with Fortney \cite{fortney2005}.
\begin{acknowledgments}
T.D.R. gratefully acknowledges support from an appointment to the NASA 
Postdoctoral Program at NASA Ames Research Center, administered by 
Oak Ridge Affiliated Universities.  L.M. thanks the Agence Nationale de la 
Recherche (ANR Project ``APOSTIC" \#11BS56002, France).  M.S.M. and 
J.J.F. acknowledge support from NASA's Planetary Atmospheres program.  
J.J.F. also acknowledges support from the NSF.  We thank W.~B. Hubbard, 
P. Muirhead, and an anonymous referee for friendly and constructive 
feedback on earlier versions of this work.
\end{acknowledgments}
%

\bibliographystyle{pnas}
%



%
\end{article}

\end{document}